\documentclass[reprint, amsmath, amssymb, aps, pre]{revtex4-1}
\usepackage{graphicx, dcolumn, bm, amsmath, amssymb, xcolor, enumitem, float}
\usepackage[spaces]{grffile}

\newcommand{\pmat}[2]{\def\arraystretch{#2}\begin{pmatrix}#1\end{pmatrix}}

\begin{document}
	\title{Three-body problem for Langevin dynamics with different temperatures}
	\author{Michael Wang}
		\email{mw3189@nyu.edu}
	\author{Alexander Y. Grosberg}
		\email{ayg1@nyu.edu}
	\affiliation{Department of Physics and Center for Soft Matter Research, New York University, 726 Broadway, New York, New York 10003, USA}
	
	\begin{abstract}
		A mixture of Brownian particles at different temperatures has been a useful model for studying the out-of-equilibrium properties of systems made up of microscopic components with differing levels of activity.  This model was previously studied analytically for two-particle interactions in the dilute limit, yielding a Boltzmann-like two-particle distribution with an effective temperature.  Like the Newtonian two and three-body problems, we ask here whether the two-particle results can be extended to three-particle interactions to get the three-particle distributions.  By considering the special solvable case of pairwise quadratic interactions, we show that, unlike the two-particle distribution, the three-particle distribution cannot in general be Boltzmann-like with an effective temperature.  We instead find that the steady state distribution of any two particles in a triplet depends on the properties of and interactions with the third particle, leading to some unexpected behaviors not present in equilibrium.
	\end{abstract}
	
	\maketitle
	
	\section{\label{sec:Introduction}Introduction}
		Common to many biological and artificial out-of-equilibrium systems is the presence of a local drive that pushes the systems away from equilibrium, that is, individual components within a system locally consume energy to generate forces and motion \cite{Marchetti et al}.  Such systems include molecular motors \cite{Astumian,Julicher et al,Sanchez et al}, artificial swimmers \cite{Palacci et al,Paxton et al,Kumar et al,Walsh et al}, bacteria \cite{Berg and Brown}, and biological tissues \cite{Bi et al, Park et al}.  One class of models frequently used to study these systems is active particles, consisting of particles self-propelled by a local force whose direction changes randomly.  It has been observed that such systems exhibit emergent, out-of-equilibrium phenomenon such as phase separation in the absence of attractive interactions \cite{Cates and Tailleur} and preferential motion in ratchet-like systems \cite{Galajda et al,Di Leonardo et al}.  
		
		Another recently studied class of models consists of mixtures of particles in contact with thermostats at different temperatures \cite{Grosberg Joanny nonequilibrium,Weber et al,Smrek Kremer phase,Smrek Kremer interface,Mura et al,Grosberg Joanny dissipation,Tanaka et al,Fogedby and Imparato,Netz,Dotsenko et al,Crisanti et al,Exartier Peliti,Falasco et al 1,Szamel,Rings et al,Falasco et al 2}.  This type of system is in some way similar to a mixture of passive and active particles, where the active particles are instead treated as having a higher temperature and hence, a higher diffusivity.  As has been shown in theory \cite{Grosberg Joanny nonequilibrium} and simulations \cite{Weber et al,Tanaka et al}, uniform mixtures of such particles become unstable for large temperature ratios and tend to phase separate into distinct regions of cold and hot particles.  This behavior is most striking in a mixture of interacting polymers at different temperatures, where even a moderate temperature ratio can lead to strong phase separation \cite{Smrek Kremer phase,Smrek Kremer interface}.  
		
		A mixture of two types of particles at temperatures $T_1\ne T_2$ was studied analytically in the dilute limit \cite{Grosberg Joanny nonequilibrium} using the steady state pair distribution of particle types $i,j$ in the mixture given by  ($k_B=1$)
		\begin{equation}
			P_{ij}(\boldsymbol{r},\boldsymbol{r}')\sim\exp\left[-\frac{1}{T_{ij}}U_{ij}(|\boldsymbol{r}-\boldsymbol{r}'|)\right],
			\label{eq:pair IJ}
		\end{equation}
		where the parameter $T_{ij}$, which we call a pairwise temperature, is 
		\begin{equation}
			T_{ij}=\frac{\gamma_jT_i+\gamma_iT_j}{\gamma_i+\gamma_j},
			\label{eq:pair TIJ}
		\end{equation}
		and $\gamma_1,\gamma_2$ are friction coefficients of the two types of particles.  While $P_{11},P_{22}$ with $T_{11}=T_1$ and $T_{22}=T_2$ are the equilibrium pair distributions for like particles, $P_{12}$ and $T_{12}$ are entirely new; in particular, they involve the transport properties $\gamma_1,\gamma_2$ of the particles, which is impossible in equilibrium systems.  Using these pair distributions, it was found that the system has non-equilibrium analogs to free energy, chemical potential, and pressure up to a second virial-like approximation \cite{Grosberg Joanny nonequilibrium}.  
		
		In analogy with the Newtonian mechanics, a natural next step from studying two-particle distributions for Langevin dynamics with different temperatures is to consider the three-particle behaviors and determine the three-particle distributions.  We ask here whether Boltzmann-like distributions similar to Eq.\ (\ref{eq:pair IJ}) can be found for three interacting particles.  As one might expect, there is a significant increase in difficulty going from the two-body to three-body problem.  By considering pairwise additive quadratic potentials, we show that even in steady state, the distribution for three particles does not take on a simple Boltzmann-like form with one effective temperature as it does for two particles.  Instead, the three-particle distribution takes on a generalized form where there are three distinct pairwise temperatures for each pair of particles.  Moreover, each pairwise temperatures depends strongly on the properties of all three particles and in addition, on the interactions between them.  This leads to some peculiar behaviors that are only present when the temperatures or activities are different.
		
	\section{The general problem}
		The usual starting point for studying a mixture of colloidal particles is a system of overdamped Langevin equations.  One approach for incorporating different levels of activity is by placing each particle in contact with thermostats of differing temperatures, which can be described by having different noise amplitudes in the Langevin equations for each particle or degree of freedom \cite{Grosberg Joanny nonequilibrium,Weber et al,Tanaka et al,Mura et al,Netz,Fogedby and Imparato,Dotsenko et al,Crisanti et al,Exartier Peliti,Falasco et al 1,Szamel,Rings et al,Falasco et al 2}.  The overdamped Langevin equations describing our three particles are
		\begin{equation}
			\gamma_i\dot{\boldsymbol{r}}_i=-\nabla_{\boldsymbol{r}_i}U(\boldsymbol{r}_1,\boldsymbol{r}_2,\boldsymbol{r}_3)+\sqrt{2T_i\gamma_i}\boldsymbol{\xi}_i,
			\label{eq:Langevin}
		\end{equation}
		where $\gamma_i$ are the friction coefficients, $U$ is some potential energy, and $\boldsymbol{\xi}_i$ are independent, unit-variance Gaussian white noises (and with independent Cartesian components).  Here and throughout the rest of the paper, we set $k_B=1$.  Note that we can reformulate this system in terms of diffusivities $D_i=T_i/\gamma_i$.  The key here is that the quantities $D_i\gamma_i$ are not equal to one temperature, that is, Einstein's relation \cite{Einstein} is violated and the system is not in equilibrium.
		
		For our present purposes, we focus on the pairwise additive interaction potential $U(\boldsymbol{r}_1,\boldsymbol{r}_2,\boldsymbol{r}_3)=U_{12}(\boldsymbol{r}_1-\boldsymbol{r}_2)+U_{23}(\boldsymbol{r}_2-\boldsymbol{r}_3)+U_{13}(\boldsymbol{r}_1-\boldsymbol{r}_3)$, much like what is typically done for the equilibrium theory of classical interacting particles \cite{Uhlenbeck Ford}.  Finally, in the context of phase separation in a mixture of $\mathcal{A}$ and $\mathcal{B}$ particles, the three particles now considered can be either $\mathcal{A}\mathcal{A}\mathcal{A}$, $\mathcal{A}\mathcal{A}\mathcal{B}$, $\mathcal{A}\mathcal{B}\mathcal{B}$, and $\mathcal{B}\mathcal{B}\mathcal{B}$ with temperatures $\{T_{\mathcal{A}},T_{\mathcal{A}},T_{\mathcal{A}}\}$, $\{T_{\mathcal{A}},T_{\mathcal{A}},T_{\mathcal{B}}\}$, $\{T_{\mathcal{A}},T_{\mathcal{B}},T_{\mathcal{B}}\}$, and $\{T_{\mathcal{B}},T_{\mathcal{B}},T_{\mathcal{B}}\}$, respectively.  To simplify the analysis, we find it easier to keep the temperatures $T_1,T_2,T_3$ general.
		
		The Langevin equation Eq.\ (\ref{eq:Langevin}) can be recast into the Fokker-Planck equation $\partial_tP=-\sum_i\nabla_{\boldsymbol{r}_i}\cdot\boldsymbol{J}_i$, where the currents $\boldsymbol{J}_i$ are
		\begin{equation}
			\boldsymbol{J}_i=-\frac{1}{\gamma_i}\left(\nabla_{\boldsymbol{r}_i}U\right)P-\frac{T_i}{\gamma_i}\nabla_{\boldsymbol{r}_i}P.
		\end{equation}
		In equilibrium with $T_1=T_2=T_3=T$, the currents $\boldsymbol{J}_i$ are zero, which gives the usual Boltzmann distribution
		\begin{equation}
			P(\boldsymbol{r}_1,\boldsymbol{r}_2,\boldsymbol{r}_3)\sim\exp\left[-\frac{1}{T}U(\boldsymbol{r}_1,\boldsymbol{r}_2,\boldsymbol{r}_3)\right].
		\end{equation}
		Out of equilibrium in steady state, the divergence of the current is still zero, that is, $\sum_i\nabla_{\boldsymbol{r}_i}\cdot\boldsymbol{J}_i=0$.  However, the currents themselves do not have to be zero as they can have nonzero curls, which makes it difficult to determine the three-particle distribution.  Based on the results of the two-particle case, a simple guess for the three-particle distribution is
		\begin{equation}
			P(\boldsymbol{r}_1,\boldsymbol{r}_2,\boldsymbol{r}_3)\sim\exp\left[-\frac{1}{T_{\textrm{eff}}}U(\boldsymbol{r}_1,\boldsymbol{r}_2,\boldsymbol{r}_3)\right],
			\label{eq:guess 1}
		\end{equation}
		for some effective temperature $T_{\textrm{eff}}$.  However, as we will see in Section \ref{sec:quadratic interactions} for the simple case of pairwise quadratic interactions, the distribution takes on the form
		\begin{equation}
			P(\boldsymbol{r}_1,\boldsymbol{r}_2,\boldsymbol{r}_3)\sim\exp\left[-\frac{U_{12}}{T_{12}}-\frac{U_{23}}{T_{23}}-\frac{U_{13}}{T_{13}}\right],
			\label{eq:guess 2}
		\end{equation}
		where $U_{ij}=U_{ij}(\boldsymbol{r}_i-\boldsymbol{r}_j)$ and $T_{ij}$, which we call ``pairwise temperatures'', depend on properties of and interactions between all three particles and not just each particle pair $i,j$.
		
	\section{\label{sec:quadratic interactions}Pairwise quadratic interactions}
		Suppose the particles are connected by ideal springs with potential energies $U_{ij}=\frac{1}{2}\kappa_{ij}(\boldsymbol{r}_i-\boldsymbol{r}_j)^2$.  It is worth mentioning that while quadratic potentials are special, they have been useful in many contexts such as membranes, proteins, and polymers \cite{Rouse,Zimm,Haliloglu et al,Helfrich,Mura et al,Osmanovic and Rabin,Eisenstecken et al,Prost and Bruinsma,Netz}.  The Langevin equations describing the system with potential energy $U=U_{12}+U_{23}+U_{13}$ are
		\begin{subequations}
			\begin{align}
				\gamma_1\dot{\boldsymbol{r}}_1&=-\kappa_{12}(\boldsymbol{r}_1-\boldsymbol{r}_2)-\kappa_{13}(\boldsymbol{r}_1-\boldsymbol{r}_3)+\sqrt{2T_1\gamma_1}\boldsymbol{\xi}_1,\label{eq:harmonic eq motion a}\\
				\gamma_2\dot{\boldsymbol{r}}_2&=-\kappa_{12}(\boldsymbol{r}_2-\boldsymbol{r}_1)-\kappa_{13}(\boldsymbol{r}_2-\boldsymbol{r}_3)+\sqrt{2T_2\gamma_2}\boldsymbol{\xi}_2,\label{eq:harmonic eq motion b}\\
				\gamma_3\dot{\boldsymbol{r}}_3&=-\kappa_{13}(\boldsymbol{r}_3-\boldsymbol{r}_1)-\kappa_{23}(\boldsymbol{r}_3-\boldsymbol{r}_2)+\sqrt{2T_3\gamma_3}\boldsymbol{\xi}_3.\label{eq:harmonic eq motion c}
			\end{align}
		\end{subequations}
	  	One can establish by direct inspection that the steady state distribution of the separations between the particles can be written as
		\begin{align}
			\begin{split}
				P(\boldsymbol{r}_1,\boldsymbol{r}_2,\boldsymbol{r}_3)&\sim\exp\left[-\frac{\kappa_{12}}{2T_{12}}(\boldsymbol{r}_1-\boldsymbol{r}_2)^2-\frac{\kappa_{23}}{2T_{23}}(\boldsymbol{r}_2-\boldsymbol{r}_3)^2\right.\\
				&\hspace{1.0in}\left.-\frac{\kappa_{13}}{2T_{13}}(\boldsymbol{r}_1-\boldsymbol{r}_3)^2\right],
			\end{split}
			\label{eq:distribution general form}
		\end{align}
		where the pairwise temperatures $T_{12},T_{23},T_{13}$ can be found from the steady state Fokker-Planck equation $\sum_i\nabla_{\boldsymbol{r}_i}\cdot\boldsymbol{J}_i=0$.  Note that for boundary conditions, we simply require the distribution to decay sufficiently fast for infinite separations such that it is normalizable.  The details of the calculation and the general expressions for $T_{ij}$ are shown in Appendix \ref{app:general solution}.  The forms of $T_{ij}$ are quite cumbersome although in equilibrium when $T_1=T_2=T_3=T$, we return to the usual Boltzmann distribution where $T_{12}=T_{23}=T_{13}=T$.  A quick glance immediately shows that the pairwise temperatures $T_{ij}$ are not equal to each other.  Since pairwise additive potentials is a simple case, this implies that the three-particle distribution cannot in general be written in a Boltzmann-like form with one effective temperature given by Eq.\ (\ref{eq:guess 1}).  In addition, the pairwise temperatures depend not only on properties of all of the particles, but also on the interactions (spring constants) between them.  This means that for more general pairwise potentials, the exponent in Eq.\ (\ref{eq:guess 2}) cannot be written as a linear combination of pair potentials $U_{ij}$ such that the coefficients or inverse pairwise temperatures $T_{ij}^{-1}$ are independent of positions.  
		
		To summarize, the simple case of pairwise additive quadratic potentials shows that the three-particle distribution does not generalize the same way as the two-particle distribution when the temperatures of the particle are different.
		
	\section{\label{sec:special cases}Simplified cases}
		Section \ref{sec:quadratic interactions} outlines a simple proof showing that the three-particle distribution for Brownian particles at different temperatures cannot be written in a Boltzmann-like form with one effective temperature.  In addition, it shows that the pairwise temperatures $T_{ij}$ for particles $i,j$ depend not only on the properties of the third particle but also the interactions with it.  Beyond the proof, and in the spirit of a more general three-body problem for Langevin dynamics, it is now interesting to study the distributions and pairwise temperatures by looking at simpler forms to see how differences in temperatures or activities of the three particles affects their behaviors.
		
		\subsection{\label{subsec:equal everything}Identical particles and springs, but different temperatures}
			Let us first consider the simplest non-equilibrium case when the temperatures $T_1,T_2,T_3$ are different while $\gamma_1=\gamma_2=\gamma_3=\gamma$ and $\kappa_{12}=\kappa_{23}=\kappa_{13}=\kappa$.  The pairwise temperatures $T_{ij}$ in Eq.\ (\ref{eq:distribution general form}) reduce to
			\begin{equation}
				T_{ij}=\frac{T_iT_j}{3}\left(\frac{1}{T_1}+\frac{1}{T_2}+\frac{1}{T_3}\right).
			\end{equation}
			There are some observations to make:
			\begin{enumerate}[leftmargin=*]
				\item[$\bullet$] The correlation of particles 1 and 2 relative to 3 is given by $\langle\boldsymbol{r}_{13}\cdot\boldsymbol{r}_{23}\rangle=\frac{T_3}{3\kappa}$, which is interestingly not controlled by temperatures $T_1,T_2$.  If we take $T_3\rightarrow0$, that is, particle 3 is no longer driven by a bath, the three-particle distribution becomes
				\begin{equation}
					P\sim\exp\left[-\frac{3\kappa}{2T_2}(\boldsymbol{r}_2-\boldsymbol{r}_3)^2-\frac{3\kappa}{2T_1}(\boldsymbol{r}_1-\boldsymbol{r}_3)^2\right].
				\end{equation}
				and of course we have $\langle\boldsymbol{r}_{13}\cdot\boldsymbol{r}_{23}\rangle=0$.  This suggests that relative to particle 3, particles 1 and 2 behave like two non-interacting Brownian particles despite there being a spring connecting the two.  In a sense, the system $\boldsymbol{r}_{13},\boldsymbol{r}_{23}$ becomes floppy.
				
				\item[$\bullet$] We may also be interested in how the distribution of a pair of particles is affected by the third.  Integrating out, say, particle 3 (the choice does not matter in this case), we obtain $P(\boldsymbol{r}_1,\boldsymbol{r}_2)\sim\exp\left[-\frac{\kappa}{2\widetilde{T}_{12}}(\boldsymbol{r}_1-\boldsymbol{r}_2)^2\right]$ where the effective pairwise temperature $\widetilde{T}_{12}$ is
				\begin{equation}
					\widetilde{T}_{12}=\frac{T_1+T_2}{3}.
				\end{equation}
				The average potential energy stored in the spring between particles 1 and 2 is $\langle U_{12}\rangle=\frac{d}{2}\widetilde{T}_{12}$, where $d$ is the spacial dimension.  Note that the distribution $P(\boldsymbol{r}_1,\boldsymbol{r}_2)$ and the average potential energy are independent of the activity (temperature) of particle 3.  This, however, is not in general true as we will see when the particles and springs are not all identical.
			\end{enumerate}
			The pairwise temperatures can be generalized to $N$ identical particles, all connected by identical springs.  Details of the calculation can be found in Appendix \ref{app:proof for N}.  The pairwise temperatures are
			\begin{equation}
				T_{ij}=\frac{T_iT_j}{N}\sum_{n=1}^{N}\frac{1}{T_n}=\frac{T_iT_j}{T_{\textrm{H}}},
			\end{equation}
			where $T_{\textrm{H}}$ is the harmonic mean of all the temperatures.  Interestingly, if we take for example $T_N=0$, particles $1,2,\dots,N-1$ appear to be noninteracting relative to particle $N$.  The potential energy stored in the spring between particles $i,j$ generalizes as
			\begin{equation}
				\langle U_{ij}\rangle=\frac{d}{2N}(T_i+T_j).
			\end{equation}
			This potential energy is independent of the temperatures of the other particles.
		
		\subsection{\label{subsec:soft stiff}Softer or stiffer spring}
			Let us now consider a slightly more complex case by softening or stiffening the spring between one pair of the particles, for example, $\kappa_{23}=\kappa_{13}=\kappa\ne\kappa_{12}$.  The pairwise temperatures are
			\begin{subequations}
				\begin{align}
					T_{12}&=\frac{\kappa_{12}A}{2T_3(\kappa+2\kappa_{12})-(T_1+T_2)(\kappa-\kappa_{12})},\\
					T_{23}&=\frac{(2\kappa+\kappa_{12})A}{3\left[3T_1(\kappa+\kappa_{12})+T_2(\kappa-\kappa_{12})\right]},\\
					T_{13}&=\frac{(2\kappa+\kappa_{12})A}{3\left[3T_2(\kappa+\kappa_{12})+T_1(\kappa-\kappa_{12})\right]},
				\end{align}
			\end{subequations}
			where
			\begin{equation}
				A=\frac{(T_1-T_2)^2(\kappa-\kappa_{12})^2}{2(2\kappa+\kappa_{12})^2}+2(T_1T_2+T_2T_3+T_1T_3)
			\end{equation}
			\begin{enumerate}[leftmargin=*]
				\item[$\bullet$] A simple limit to check is $\kappa_{12}\rightarrow\infty$.  We expect particles 1 and 2 effectively merge and the system to reduce to two particles.  Taking the limit, we see that the three-particle distribution becomes
				\begin{equation}
					P(\boldsymbol{r}_1,\boldsymbol{r}_2,\boldsymbol{r}_3)\sim\delta(\boldsymbol{r}_1-\boldsymbol{r}_2)\exp\left[-\frac{\kappa}{T_{\{12\}3}}(\boldsymbol{r}_1-\boldsymbol{r}_3)^2\right],
					\label{eq:distribution k12 infinity}
				\end{equation}
				where $T_{\{12\}3}=\frac{T_1+T_2+4T_3}{6}$, which denotes the pairwise temperature of the combined particle \{12\} and particle 3.  Eq.\ (\ref{eq:distribution k12 infinity}) can correctly be interpreted as the distribution of two particles with properties $T_3,\gamma$ and $\frac{T_1+T_2}{2},2\gamma$ connected by a spring with combined stiffness $2\kappa$.  This result does have a simple generalization for arbitrary springs and friction coefficients, which we discuss in Section \ref{subsec:remark}.
				
				\item[$\bullet$] The correlations of particles 1 and 2 relative to 3 in this case become
				\begin{equation}
					\langle\boldsymbol{r}_{13}\cdot\boldsymbol{r}_{23}\rangle=\frac{T_3-T_3^*}{3\kappa},
				\end{equation}
				where $T_3^*=\frac{(T_1+T_2)(\kappa-\kappa_{12})}{2(\kappa+2\kappa_{12})}$.  Here, we see an effect that is not present in equilibrium.  In equilibrium when $T_1=T_2=T_3=T$, we have $\langle\boldsymbol{r}_{13}\cdot\boldsymbol{r}_{23}\rangle=\frac{\kappa_{12}T}{\kappa(\kappa+2\kappa_{12})}$, which is positive for any $\kappa_{12}>0$.  This is of course due to the spring connecting particles 1 and 2.  When the temperatures are not equal and $\kappa_{12}<\kappa$, however, we see that $\langle\boldsymbol{r}_{13}\cdot\boldsymbol{r}_{23}\rangle$ can be negative and even zero.  In particular, $\langle\boldsymbol{r}_{13}\cdot\boldsymbol{r}_{23}\rangle<0$ when $T_3<T_3^*$ and particles 1 and 2 appear to be repulsive relative to 3 while $\langle\boldsymbol{r}_{13}\cdot\boldsymbol{r}_{23}\rangle=0$ when $T_3=T_3^*$ and instead they appear non-interacting even though there is a spring between them.  In terms of the distribution and the coefficient $T_{12}^{-1}$ of the pair potential $U_{12}$, these correspond to $T_{12}^{-1}<0$ when $T_3<T_3^*$ and $T_{12}^{-1}=0$ when $T_3=T_3^*$.  If $\kappa_{12}>\kappa$, there are similar sign changes in $T_{23}^{-1}$ and $\langle\boldsymbol{r}_{21}\cdot\boldsymbol{r}_{31}\rangle$ at $T_1=\frac{T_2(\kappa_{12}-\kappa)}{3(\kappa_{12}+\kappa)}$ and $T_{13}^{-1}$ and $\langle\boldsymbol{r}_{12}\cdot\boldsymbol{r}_{32}\rangle$ at $T_2=\frac{T_1(\kappa_{12}-\kappa)}{3(\kappa_{12}+\kappa)}$.  Note that equilibrium, we simply have $T_{12}=T_{23}=T_{13}=T$ irrespective of the choice of $\kappa_{12}$.
				
				\item[$\bullet$] Like in Section \ref{subsec:equal everything}, we may be interested in how a pair of particles is affected by the third.  Here, there are two choices of particles to integrate out.  Integrating out particle 3, the particle across from the spring between particles 1 and 2, we have $P(\boldsymbol{r}_1,\boldsymbol{r}_2)\sim\exp\left[-\frac{\kappa_{12}}{2\widetilde{T}_{12}}(\boldsymbol{r}_1-\boldsymbol{r}_2)^2\right]$ with
				\begin{equation}
					\widetilde{T}_{12}=\frac{\kappa_{12}(T_1+T_2)}{\kappa+2\kappa_{12}}.
				\end{equation}
				If we instead integrated out, say, particle 2, we have $P(\boldsymbol{r}_1,\boldsymbol{r}_3)\sim\exp\left[-\frac{\kappa}{2\widetilde{T}_{13}}(\boldsymbol{r}_1-\boldsymbol{r}_3)^2\right]$ with
				\begin{align}
					\begin{split}
						\widetilde{T}_{13}&=\frac{1}{6(2\kappa+\kappa_{12})(\kappa+2\kappa_{12})}\Big[T_1(7\kappa^2+10\kappa\kappa_{12}+\kappa_{12}^2)\\
						&\hspace{0.2in}+2T_3(2\kappa+\kappa_{12})(\kappa+2\kappa_{12})+T_2(\kappa-\kappa_{12})^2\Big].
					\end{split}
					\label{eq:tilde T13 k12}
				\end{align}
				The average potential energy stored in the spring between particles 1 and 2 is $\langle U_{12}\rangle=\frac{d}{2}\widetilde{T}_{12}$.  This energy depends on $\kappa$, the interaction with particle 3, but is still independent of the temperature $T_3$.  In other words, the activity of the integrated-out particle does not affect the remaining two.  This, however, is not the case for the average energy $\langle U_{13}\rangle=\frac{d}{2}\widetilde{T}_{13}$ stored in the spring between particles 1 and 3, as is evident by the appearance of $T_2$ in $\widetilde{T}_{13}$.  In equilibrium, $\widetilde{T}_{13}=\frac{(\kappa+\kappa_{12})T}{\kappa+2\kappa_{12}}$.
			\end{enumerate}
		
		\subsection{Particles with different mobilities}
			In equilibrium, the mobilities or transport properties of particles cannot enter into the Boltzmann distribution while out of equilibrium, they can.  Suppose that $\gamma_1=\gamma_2=\gamma\ne\gamma_3$ while keeping the springs identical.  This is the simplest case where the mobilities do not automatically drop out of the steady state Fokker-Planck equation $\sum_i\nabla_{\boldsymbol{r}_i}\cdot\boldsymbol{J}_i=0$.  The pairwise temperatures are
			\begin{subequations}
				\begin{align}
					T_{12}&=\frac{A}{6T_3\gamma-(T_1+T_2)(\gamma-\gamma_3)},\\
					T_{23}&=\frac{(\gamma+2\gamma_3)A}{\big[T_1(\gamma+5\gamma_3)+T_2(\gamma-\gamma_3)\big](2\gamma+\gamma_3)},\\
					T_{13}&=\frac{(\gamma+2\gamma_3)A}{\big[T_2(\gamma+5\gamma_3)+T_1(\gamma-\gamma_3)\big](2\gamma+\gamma_3)},
				\end{align}
			\end{subequations}
			where
			\begin{align}
				\begin{split}
					A&=\frac{(T_1-T_2)^2\gamma_3(\gamma-\gamma_3)^2}{2(\gamma+2\gamma_3)^2}\\
					&\hspace{0.4in}+2(T_1T_2\gamma_3+T_2T_3\gamma+T_1T_3\gamma).
				\end{split}
			\end{align}
			Just as the case with a softer or stiffer spring, if we integrate out particle 3, we find
			\begin{equation}
				\widetilde{T}_{12}=\frac{T_1+T_2}{3}
			\end{equation}
			If we instead integrate out particle 2, we obtain 
			\begin{align}
				\begin{split}
					\widetilde{T}_{13}&=\frac{1}{6(2\gamma+\gamma_3)(\gamma+2\gamma_3)}\Big[T_1(\gamma^2+10\gamma\gamma_3+7\gamma_3^2)\\
					&\hspace{0.7in}+6T_3\gamma(\gamma+2\gamma_3)+T_2(\gamma-\gamma_3)^2\Big].
				\end{split}
			\end{align}
			Just as the case with softening or stiffening one of the springs in Section \ref{subsec:soft stiff}, we also observe similar sign changes in $T_{ij}^{-1}$ and $\langle\boldsymbol{r}_{ik}\cdot\boldsymbol{r}_{jk}\rangle$ depending on the choices of $\gamma_3$ and the temperatures.  In addition, the average energy stored in the spring between particles 1 and 3 $\langle U_{13}\rangle=\frac{d}{2}\widetilde{T}_{13}$ also depends on the temperature $T_2$ of the integrated out particle.  What is different here is that in equilibrium, the friction coefficients play no role in the distribution and behaviors of the three particles; in particular, $\widetilde{T}_{12}=\widetilde{T}_{23}=\widetilde{T}_{13}=\frac{2T}{3}$ are completely independent of $\gamma$'s.  Only when system is out of equilibrium and the temperatures are different do the mobilities have a significant effect.
			
		\subsection{\label{subsec:remark}Remark on the general case}
			The forms of $T_{ij}$ for the general case of different spring constants and friction coefficients are quite cumbersome (Appendix \ref{app:general solution}).  There is a simple case.  As mentioned in Section \ref{subsec:soft stiff}, taking $\kappa_{12}\rightarrow\infty$ corresponds to effectively merging particles 1 and 2, which reduces the system to a combined particle \{12\} and particle 3.  The properties of particle 3 are simply $T_3,\gamma_3$.  The combined particle \{12\} will have a total friction coefficient $\gamma_{\{12\}}=\gamma_1+\gamma_2$.  The total drive on particle \{12\} is $\xi_{\{12\}}=\sqrt{2T_1\gamma_1}\xi_1+\sqrt{2T_2\gamma_2}\xi_2$ with correlations $\langle\xi_{\{12\}}(t)\xi_{\{12\}}(t')\rangle=2T_{\{12\}}\gamma_{\{12\}}\delta(t-t')$, where the effective temperature of \{12\} is $T_{\{12\}}=\frac{\gamma_1T_2+\gamma_2T_2}{\gamma_1+\gamma_2}$.  Following the same notation as in Section \ref{subsec:soft stiff}, the pairwise temperature $T_{\{12\}3}$ of the combined particle \{12\} and particle 3 in the limit $\kappa_{12}\rightarrow\infty$ is the mobility weighted average (Eq.\ (\ref{eq:pair TIJ}))
			\begin{align}
				\begin{split}
					T_{\{12\}3}&=\frac{\gamma_3T_{\{12\}}+\gamma_{\{12\}}T_3}{\gamma_3+\gamma_{\{12\}}}\\
					&=\frac{\gamma_3\frac{\gamma_1T_1+\gamma_2T_2}{\gamma_1+\gamma_2}+(\gamma_1+\gamma_2)T_3}{\gamma_3+(\gamma_1+\gamma_2)}.
				\end{split}
			\end{align}
			Since the springs from particle 3 to particles 1 and 2 are in parallel, the total spring constant between particle \{12\} and particle 3 is $\kappa_{13}+\kappa_{23}$.  Note that when the friction coefficients are all equal, we get back $T_{\{12\}3}=\frac{T_1+T_2+4T_3}{6}$ in Section \ref{subsec:soft stiff}.
		
	\section{Underdamped particles}
		As we saw in Section \ref{sec:quadratic interactions} for pairwise quadratic interactions, including a new degree of freedom, a third particle, leads to a complicated distribution that cannot be written in a generalized Boltzmann-like form where the pairwise temperatures depend only on pair properties of the particles (Eq.\ (\ref{eq:guess 2})).  Instead, the pairwise temperatures must depend on the properties of all the particles, and in addition on the springs between them.  Because of that, it is interesting to look at another way of including additional degrees of freedom by considering underdamped particles, where there are momenta in addition to positions.
		
		\subsection{\label{subsec:two underdamped}Two underdamped particles}
			Consider the case of two underdamped particles, where we now have four degrees of freedom: two positions and two momenta.  For simplicity, suppose both particles have the same mass $m$ and friction coefficient $\gamma$, but different temperatures $T_1,T_2$.  The Langevin equations are
			\begin{subequations}
				\begin{align}
					&\dot{\boldsymbol{r}}=\frac{1}{m}(\boldsymbol{p}_1-\boldsymbol{p}_2),\\
					&\dot{\boldsymbol{p}}_1=-\frac{\gamma}{m}\boldsymbol{p}_1-\kappa \boldsymbol{r}+\sqrt{2T_1\gamma}\boldsymbol{\xi}_1,\\
					&\dot{\boldsymbol{p}}_2=-\frac{\gamma}{m}\boldsymbol{p}_2+\kappa \boldsymbol{r}+\sqrt{2T_2\gamma}\boldsymbol{\xi}_2,
				\end{align}
			\end{subequations}
			where $\boldsymbol{p}_i=m\dot{\boldsymbol{r}}_i$ and $\boldsymbol{r}=\boldsymbol{r}_1-\boldsymbol{r}_2$.  We find the steady-state probability distribution
			\begin{align}
				\begin{split}
					&P(\boldsymbol{r},\boldsymbol{p}_1,\boldsymbol{p}_2)\sim\exp\left[{-\frac{1}{2}}\Big(\beta_{p_1p_1}\boldsymbol{p}_1^2+\beta_{p_2p_2}\boldsymbol{p}_2^2+\beta_{rr}\boldsymbol{r}^2\right.\\
					&\hspace{0.4in}\left.\vphantom{\frac{1}{2}}+2\beta_{p_1p_2}\boldsymbol{p}_1\cdot\boldsymbol{p}_2+2\beta_{p_1r}\boldsymbol{p}_1\cdot\boldsymbol{r}+2\beta_{p_2r}\boldsymbol{p}_2\cdot\boldsymbol{r}\Big)\right],
				\end{split}
				\label{eq:underdamped distribution}
			\end{align}
			where the off-diagonal coefficients $\beta_{p_1p_2},\beta_{p_1r},\beta_{p_2r}$ are nonzero when $T_1\ne T_2$.  Such a distribution with nonzero cross terms has been reported in similar systems \cite{Mura et al,Netz}.  
			
			Of course in equilibrium when $T_1=T_2=T$, we obtain the usual Boltzmann distribution $P\sim\exp\left[-\frac{1}{T}\left(\frac{\boldsymbol{p}_1^2}{2m}+\frac{\boldsymbol{p}_2^2}{2m}+\frac{1}{2}\kappa\boldsymbol{r}^2\right)\right]$.  When the temperatures are different, however, we see that the distribution does not generalize the same way as it did for overdamped particles, that is, the distributions indicates correlations between momenta and positions.
		
		\subsection{Kinetic energies and potential energies}
			For three or more particles, the distribution takes on a form similar to Eq.\ (\ref{eq:underdamped distribution}) for $\boldsymbol{p}_1,\boldsymbol{p}_2,\boldsymbol{p}_3,\boldsymbol{r}_1-\boldsymbol{r}_2,\boldsymbol{r}_2-\boldsymbol{r}_3,\boldsymbol{r}_1-\boldsymbol{r}_3$.  The expressions, however, are significantly more cumbersome.  We can still compute the average kinetic and potential energies.  The case of $N$ identical particles and springs can be found in Appendix \ref{subapp:N underdamped particles}. 
			
			For $N$ identical particles all connected by identical springs, the average kinetic energy $\langle K_i\rangle$ of the $i^{\textmd{th}}$ particle is
			\begin{equation}
				\langle K_i\rangle=\frac{d}{2}T_i-\frac{dm\kappa}{2\gamma^2+Nm\kappa}\left(T_i-T_{\textrm{avg}}\right),
				\label{eq:N particle KE}
			\end{equation}
			where $T_{\textrm{avg}}=\frac{1}{N}\sum_{n=1}^{N}T_n$ is the average temperature.  The potential energy stored between particles $i,j$ is
			\begin{equation}
				\langle U_{ij}\rangle=\frac{d}{2N}(T_i+T_j).
			\end{equation}
			This result is independent of mass and the other temperatures, similar to the overdamped case.  However, if we change the interactions between the particles, for example taking $\kappa_{23}=\kappa_{13}=\kappa\ne\kappa_{12}$ as before, the potential energies show a dependence on the mass.  The results can be found in Appendix {\ref{subapp:underdamped one spring}}.
		
	\section{\label{sec:discussion and conclusion}Discussion and Conclusion}		
		For the two-body problem for Langevin dynamics with different temperatures, the two-particle distribution can be written in a Boltzmann-like form with an effective temperature (Eqs. (\ref{eq:pair IJ}) and (\ref{eq:pair TIJ})) \cite{Grosberg Joanny nonequilibrium}.  For the three-body case, however, we showed that such a Boltzmann-like form with one effective temperature is not possible and instead, at least for the case of pairwise additive quadratic potentials, the steady-state three-particle distribution acquires the form
		\begin{equation}
			P(\boldsymbol{r}_1,\boldsymbol{r}_2,\boldsymbol{r}_3)\sim\exp\left[-\frac{U_{12}}{T_{12}}-\frac{U_{23}}{T_{23}}-\frac{U_{13}}{T_{13}}\right],
			\label{eq:guess 2 conclusion}
		\end{equation}
		with three distinct $T_{ij}$ which we call the pairwise temperatures.  
		
		In the two-particle case, the effective temperature $T_{12}$ depends on $T_1,T_2$ and $\gamma_1,\gamma_2$, but not on $\kappa_{12}$.  By contrast, in the three-particle case, each pairwise temperature, for example $T_{12}$, depends not only on the properties of particles 1 and 2 ($T_1,T_2$ and $\gamma_1,\gamma_2$) and their interaction ($\kappa_{12}$), but also on the properties of particle 3 ($T_3,\gamma_3$) and interactions with it ($\kappa_{13},\kappa_{23}$).  The dependence of each pairwise temperature on the spring constants suggests that the three-particle distribution cannot be generalized in a simple way to arbitrary pairwise potentials.  In other words, the exponent will not be a linear combination of pair potentials with coefficients that are independent of particle positions and interactions.  Similar difficulties and complexities arise in systems of active (self-propelled) particles.  In particular, the distributions in such systems can only be written approximately with position-dependent effective temperatures/diffusivities or effective interactions \cite{Maggi et al,Farage et al,Fodor et al}.  Although pairwise additive potentials are a special case, we should point out that for more general three-body potentials (not necessarily pairwise additive), there is no reason to expect the distributions to be simpler.
		
		The dependence of each pairwise temperature $T_{ij}$ on the properties of all three particles and interactions has some unexpected consequences on the distribution and behavior of the three particles.  In particular, the inverse of these pairwise temperature $T_{ij}^{-1}$, which are the coefficients of the pair potentials $U_{ij}$ in Eq.\ (\ref{eq:guess 2 conclusion}), can change sign and even be zero depending on the choice of temperatures $T_i$.  For example, even when there is no spring between particles 1 and 2 ($\kappa_{12}=0$) while the two other springs are the same ($\kappa_{13}=\kappa_{23}=\kappa$), the correlation
		\begin{equation}
			\langle\boldsymbol{r}_{13}\cdot\boldsymbol{r}_{23}\rangle=\frac{1}{3\kappa}\left[T_3-\frac{T_1+T_2}{2}\right]
		\end{equation} 
		can be either positive or negative depending on the relation between the temperatures.  If particle 3 is ``hot'' ($T_3>(T_1+T_2)/2$), the correlation is positive meaning that $\boldsymbol{r}_{13},\boldsymbol{r}_{23}$ roughly point in the same direction and particles 1 and 2 appear attractive.  If particle 3 is cold ($T_3<(T_1+T_2)/2$), the correlation is negative meaning $\boldsymbol{r}_{13},\boldsymbol{r}_{23}$ roughly point in opposite directions and particles 1 and 2 appear repulsive.  Note that in equilibrium, the correlation is zero since there is no spring between particle 1 and 2.
		
		In addition to the sign changes, we observe that every particle of the triplet affects the mutual behavior of the other two.  For instance, in a simple case when $\kappa_{12}=\kappa_{23}=\kappa_{13}=\kappa$ and $T_3\gg T_1,T_2$, that is particle 3 is much more active than 1 and 2, we find
		\begin{equation}
			\langle(\boldsymbol{r}_1-\boldsymbol{r}_2)^2\rangle\approx\frac{dT_3(\gamma_1-\gamma_2)^2}{12(\gamma_1+\gamma_2+\gamma_3)(\gamma_1\gamma_2+\gamma_2\gamma_3+\gamma_1\gamma_3)}.
		\end{equation}
		This is not only controlled by the temperature $T_3$ of particle 3, but also by its friction $\gamma_3$.  Note that more generally, when $T_3\gg T_1,T_2$, we have $\langle(\boldsymbol{r}_1-\boldsymbol{r}_2)^2\rangle\propto T_3\left(\frac{\gamma_1}{\kappa_{13}}-\frac{\gamma_2}{\kappa_{23}}\right)^2$.  Thus, the activity of particle 3 will affect 1 and 2 if their relaxation times $\frac{\gamma_1}{\kappa_{13}}$ and $\frac{\gamma_2}{\kappa_{23}}$ are not equal, that is, they interact differently with 3.  This behavior has some connection with allostery-inspired mechanical networks \cite{Rocks et al}, where the motion or activity of nodes in one part of a mechanical network can affect the response in another part differently depending on how one removes bonds (analogous to changing the interaction strengths between nodes).

		In Newtonian mechanics, where one would like to write down closed-form expressions for the trajectories of the interacting bodies, there is a dramatic increase in difficulty going from the two to three-body problem.  While the two-body problem can be solved by simply transforming to the center of mass and relative separation coordinates, the three-body problem remains unsolved except for special arrangements and potentials.  For Langevin dynamics, in which we would like to determine the distribution of the bodies, there is also a similar increase in difficulty and richness.  In the two-body case for overdamped particles, the distribution for any potential is Boltzmann-like with an effective temperature \cite{Grosberg Joanny nonequilibrium}.  As we showed and discussed for the special case of pairwise additive quadratic potentials, the three-particle distribution cannot be written with an effective temperature and cannot be obtained for general, non-pairwise interactions when the temperatures of the three particles are different.
		
	\section*{Acknowledgments}
		This work was supported primarily by the MRSEC Program of the National Science Foundation under Award Number DMR-1420073.  We thank J.-F. Joanny, W. Srinin, and M. L. Manning for interesting discussions and insightful comments.
		
	\appendix
	\section{\label{app:general solution}Arbitrary spring constants and friction coefficients}
		Since the system described by Eqs.\ (\ref{eq:harmonic eq motion a}), (\ref{eq:harmonic eq motion b}), and (\ref{eq:harmonic eq motion c}) is translationally invariant, it is useful to consider the relative separations defined by $\boldsymbol{r}_{13}=\boldsymbol{r}_1-\boldsymbol{r}_3$ and $\boldsymbol{r}_{23}=\boldsymbol{r}_2-\boldsymbol{r}_3$.  In these variables, we have
		\begin{subequations}
			\begin{align}
				\begin{split}
				    \dot{\boldsymbol{r}}_{13}&=-\left(\frac{\kappa_{12}}{\gamma_1}+\frac{\kappa_{13}}{\gamma_{13}}\right)\boldsymbol{r}_{13}-\left(\frac{\kappa_{23}}{\gamma_3}-\frac{\kappa_{12}}{\gamma_1}\right)\boldsymbol{r}_{23}\\
					 &\hspace{1in}+\sqrt{\frac{2T_1}{\gamma_1}}\boldsymbol{\xi}_1-\sqrt{\frac{2T_3}{\gamma_3}}\boldsymbol{\xi}_3,
				\end{split}
				\\
				\begin{split}
					\dot{\boldsymbol{r}}_{23}&=-\left(\frac{\kappa_{13}}{\gamma_3}-\frac{\kappa_{12}}{\gamma_2}\right)\boldsymbol{r}_{13}-\left(\frac{\kappa_{12}}{\gamma_2}+\frac{\kappa_{23}}{\gamma_{23}}\right)\boldsymbol{r}_{23}\\
					&\hspace{1in}+\sqrt{\frac{2T_2}{\gamma_2}}\boldsymbol{\xi}_2-\sqrt{\frac{2T_3}{\gamma_3}}\boldsymbol{\xi}_3,
				\end{split}
			\end{align}
		\end{subequations}
		where $\gamma_{i3}=\gamma_i\gamma_3/(\gamma_i+\gamma_3)$.  Computing the covariance matrix $C_{ij}=\langle\boldsymbol{r}_{i3}\boldsymbol{r}_{j3}\rangle$, where $\langle\rangle$ indicates ensemble average, we can write the steady state distribution as
		\begin{equation}
			P(\boldsymbol{r}_{13},\boldsymbol{r}_{23})\sim\exp\left[-\frac{1}{2}R^TC^{-1}R\right],
		\end{equation}
		where $R^T=\pmat{\boldsymbol{r}_{13}&\boldsymbol{r}_{23}}{1}$.  After some algebra, we find the general distribution in the main text (Eq.\ (\ref{eq:distribution general form})) given by
		\begin{align}
			\begin{split}
				P(\boldsymbol{r}_1,\boldsymbol{r}_2,\boldsymbol{r}_3)&\sim\exp\left[-\frac{\kappa_{12}}{2T_{12}}(\boldsymbol{r}_1-\boldsymbol{r}_2)^2-\frac{\kappa_{23}}{2T_{23}}(\boldsymbol{r}_2-\boldsymbol{r}_3)^2\right.\\
				&\hspace{1.0in}\left.-\frac{\kappa_{13}}{2T_{13}}(\boldsymbol{r}_1-\boldsymbol{r}_3)^2\right].
			\end{split}
		\end{align}
		The pairwise temperatures $T_{12},T_{23},T_{13}$ are given by
		\begin{widetext}
		 	\begin{subequations}
			 	\begin{align}
				 	\begin{split}
				 	 	\frac{\kappa_{12}}{T_{12}}&=\frac{1}{A}\Big[T_1\gamma_1(\gamma_3\kappa_{12}+(\gamma_2+\gamma_3)\kappa_{23})(\gamma_3\kappa_{12}-\gamma_2\kappa_{13})+T_2\gamma_2\big(\gamma_3\kappa_{12}+(\gamma_1+\gamma_3)\kappa_{13}\big)(\gamma_3\kappa_{12}-\gamma_1\kappa_{23})\\
				 	 	&\hspace{0.35in}+T_3\gamma_3\Big(\big(\gamma_1\kappa_{23}+(\gamma_1+\gamma_2)\kappa_{12}\big)\big(\gamma_2\kappa_{13}+(\gamma_1+\gamma_2)\kappa_{12}\big)+\frac{\gamma_1\gamma_2}{\gamma_3}(\gamma_1+\gamma_2+\gamma_3)(\kappa_{12}\kappa_{13}+\kappa_{12}\kappa_{23}+\kappa_{13}\kappa_{23})\Big)\Big],
				 	 	\label{eq:T12 general}
				 	\end{split}
				 	\\
				 	\begin{split}
				 	 	\frac{\kappa_{23}}{T_{23}}&=\frac{1}{A}\Big[T_3\gamma_3\big(\gamma_1\kappa_{23}+(\gamma_1+\gamma_2)\kappa_{12}\big)(\gamma_1\kappa_{23}-\gamma_2\kappa_{13})+T_2\gamma_2\big(\gamma_1\kappa_{23}+(\gamma_1+\gamma_3)\kappa_{13}\big)(\gamma_1\kappa_{23}-\gamma_3\kappa_{12})\\
						&\hspace{0.35in}+T_1\gamma_1\Big(\big(\gamma_3\kappa_{12}+(\gamma_2+\gamma_3)\kappa_{23}\big)\big(\gamma_2\kappa_{13}+(\gamma_2+\gamma_3)\kappa_{23}\big)+\frac{\gamma_2\gamma_3}{\gamma_1}(\gamma_1+\gamma_2+\gamma_3)(\kappa_{12}\kappa_{13}+\kappa_{12}\kappa_{23}+\kappa_{13}\kappa_{23})\Big)\Big],
						\label{eq:T23 general}
				 	\end{split}
				 	\\
				 	\begin{split}
				 	 	\frac{\kappa_{13}}{T_{13}}&=\frac{1}{A}\Big[T_1\gamma_1\big(\gamma_2\kappa_{13}+(\gamma_2+\gamma_3)\kappa_{23}\big)(\gamma_2\kappa_{13}-\gamma_3\kappa_{12})+T_3\gamma_3\big(\gamma_2\kappa_{13}+(\gamma_1+\gamma_2)\kappa_{12}\big)(\gamma_2\kappa_{13}-\gamma_1\kappa_{23})\\
						&\hspace{0.35in}+T_2\gamma_2\Big(\big(\gamma_1\kappa_{23}+(\gamma_1+\gamma_3)\kappa_{13}\big)\big(\gamma_3\kappa_{12}+(\gamma_1+\gamma_3)\kappa_{13}\big)+\frac{\gamma_1\gamma_3}{\gamma_2}(\gamma_1+\gamma_2+\gamma_3)(\kappa_{12}\kappa_{13}+\kappa_{12}\kappa_{23}+\kappa_{13}\kappa_{23})\Big)\Big],
						\label{eq:T13 general}
				 	\end{split}
			 	\end{align}
		 	\end{subequations}
		 	where
		 	\begin{align}
		 		\begin{split}
			 		A&=\frac{\gamma_1\gamma_2\gamma_3\big[\gamma_1\kappa_{23}(T_2-T_3)+\gamma_2\kappa_{13}(T_3-T_1)+\gamma_3\kappa_{12}(T_1-T_2)\big]^2}{\gamma_1\kappa_{23}(\gamma_2+\gamma_3)+\gamma_2\kappa_{13}(\gamma_1+\gamma_3)+\gamma_3\kappa_{12}(\gamma_1+\gamma_2)}\\
			 		&\hspace{0.8in}+(\gamma_1T_2T_3+\gamma_2T_1T_3+\gamma_3T_1T_2)\big[\gamma_1\kappa_{23}(\gamma_2+\gamma_3)+\gamma_2\kappa_{13}(\gamma_1+\gamma_3)+\gamma_3\kappa_{12}(\gamma_1+\gamma_2)\big].
		 		\end{split}
		 	\end{align}
		 	If we integrate out, say, the third particle, the probability distribution becomes
		 	\begin{equation}
			 	P(\boldsymbol{r}_1,\boldsymbol{r}_2)\sim\exp\left[-\left(\frac{1}{T_{12}}+\frac{\kappa_{13}\kappa_{23}}{\kappa_{12}(\kappa_{13}T_{23}+\kappa_{23}T_{13})}\right)\frac{1}{2}\kappa_{12}(\boldsymbol{r}_1-\boldsymbol{r}_2)^2\right]\sim\exp\left[-\frac{\kappa_{12}}{2\widetilde{T}_{12}}(\boldsymbol{r}_1-\boldsymbol{r}_2)^2\right],
		 	\end{equation}
		 	where
		 	\begin{align}
			 	\begin{split}
				 	\widetilde{T}_{12}&=\frac{2}{(\gamma_1+\gamma_2+\gamma_3)(\kappa_{12}\kappa_{13}+\kappa_{12}\kappa_{23}+\kappa_{13}\kappa_{23})\big(\gamma_1\kappa_{23}(\gamma_2+\gamma_3)+\gamma_2\kappa_{13}(\gamma_1+\gamma_3)+\gamma_3\kappa_{12}(\gamma_1+\gamma_2)\big)}\\
				 	&\hspace{0.3in}\times\Big[T_1\gamma_1\Big(\big(\gamma_2\kappa_{13}+(\gamma_2+\gamma_3)\kappa_{23}\big)^2+\frac{\gamma_2\gamma_3}{\gamma_1}(\gamma_1+\gamma_2+\gamma_3)(\kappa_{12}\kappa_{23}+\kappa_{12}\kappa_{13}+\kappa_{13}\kappa_{23})\Big)\\
				 	&\hspace{0.6in}+T_2\gamma_2\Big(\big(\gamma_1\kappa_{23}+(\gamma_1+\gamma_3)\kappa_{13}\big)^2+\frac{\gamma_1\gamma_3}{\gamma_2}(\gamma_1+\gamma_2+\gamma_3)(\kappa_{12}\kappa_{23}+\kappa_{12}\kappa_{13}+\kappa_{13}\kappa_{23})\Big)\\
				 	&\hspace{0.6in}+T_3\gamma_3(\gamma_2\kappa_{13}-\gamma_1\kappa_{23})^2\Big].
				 	\label{eq:tilde T12 general}
			 	\end{split}
		 	\end{align}
		\end{widetext}
		Note that we can obtain $\widetilde{T}_{23},\widetilde{T}_{13}$ by simply relabeling the particles.  The average potential energy stored in the spring between particles 1 and 2 is $U_{12}=\frac{d}{2}\widetilde{T}_{12}$.  Interestingly, the interaction between particles 1 and 2 depends on the temperature of particle 3 when $\kappa_{23}/\gamma_2\ne\kappa_{13}/\gamma_1$, that is, the relaxation times of particles 1 and 2 relative to 3 are not equal.
	
	\section{\label{app:proof for N}$N$ particle distribution with identical mobilities and springs}
		We here outline the results needed for the $N$ particle result.  The Langevin equations are
		\begin{equation}
			\gamma\dot{\boldsymbol{r}}_i=-\kappa\sum_{i<j}(\boldsymbol{r}_i-\boldsymbol{r}_j)+\sqrt{2T_i\gamma}\boldsymbol{\xi}_i.
		\end{equation}
		Relative to the $N$th particle, the Langevin equations for $\boldsymbol{r}_{iN}=\boldsymbol{r}_i-\boldsymbol{r}_N$ ($i=1,2,\dots,N-1$) are
		\begin{equation}
			\gamma\dot{\boldsymbol{r}}_{iN}=-N\kappa\dot{\boldsymbol{r}}_{iN}+\sqrt{2T_i\gamma}\boldsymbol{\xi}_i-\sqrt{2T_N\gamma}\boldsymbol{\xi}_N.
		\end{equation}
		The covariance matrix can easily be computed as
		\begin{equation}
			C_{ij}=\langle \boldsymbol{r}_{iN}\boldsymbol{r}_{jN}\rangle=\frac{\sqrt{T_iT_j}\delta_{ij}+T_N}{N\kappa}\boldsymbol{I}.
		\end{equation}
		The inverse of the covariance matrix is given by
		\begin{equation}
			(C^{-1})_{ij}=\kappa T_H\left(\frac{\delta_{ij}}{\sqrt{T_iT_j}}\sum_{n=1}^{N}\frac{1}{T_n}-\frac{1}{T_{i}T_j}\right)\boldsymbol{I},
		\end{equation}
		where $T_H=N\left(\sum_{n=1}^{N}\frac{1}{T_n}\right)^{-1}$ is the harmonic average of the temperatures.  It is easy to check that $CC^{-1}=C^{-1}C=I$.  After some algebra, we have
		\begin{equation}
			\sum_{i,j}^{N-1}\boldsymbol{r}_{iN}(C^{-1})_{ij}\boldsymbol{r}_{jN}=\sum_{i<j}\frac{1}{T_{ij}}(\boldsymbol{r}_i-\boldsymbol{r}_j)^2,
		\end{equation}
		where
		\begin{equation}
			T_{ij}=\frac{T_iT_j}{N}\sum_{n=1}^{N}\frac{1}{T_n}=\frac{T_iT_j}{T_H}.
		\end{equation}
		
	\section{\label{app:underdamped}Underdamped particles}
		\subsection{\label{subapp:two particles}Two particles}
			As discussed in Section \ref{subapp:two particles}, the distribution of momenta and separation between two underdamped particles at different temperatures can be written as
			\begin{align}
				\begin{split}
					&P(\boldsymbol{r},\boldsymbol{p}_1,\boldsymbol{p}_2)\sim\exp\left[{-\frac{1}{2}}\Big(\beta_{p_1p_1}\boldsymbol{p}_1^2+\beta_{p_2p_2}\boldsymbol{p}_2^2+\beta_{rr}\boldsymbol{r}^2\right.\\
					&\hspace{0.4in}\left.\vphantom{\frac{1}{2}}+2\beta_{p_1p_2}\boldsymbol{p}_1\cdot\boldsymbol{p}_2+2\beta_{p_1r}\boldsymbol{p}_1\cdot\boldsymbol{r}+2\beta_{p_2r}\boldsymbol{p}_2\cdot\boldsymbol{r}\Big)\right].
				\end{split}
			\end{align}
			The coefficients are given by
			\begin{subequations}
				\begin{align}
					\begin{split}
						A\beta_{p_1p_1}&=\frac{2A}{m(T_1+T_2)}-2\gamma^2(\gamma^2+m\kappa)T_2(T_1-T_2),
					\end{split}
					\\
					\begin{split}
						A\beta_{p_2p_2}&=\frac{2A}{m(T_1+T_2)}+2\gamma^2(\gamma^2+m\kappa)T_1(T_1-T_2),
					\end{split}
					\\
					\begin{split}
						A\beta_{rr}&=\frac{2\kappa A}{(T_1+T_2)}+2m^2\gamma^2\kappa^2(T_1-T_2)^2,
					\end{split}
					\\
					\begin{split}
						A\beta_{p_1p_2}&=m\gamma^2\kappa(T_1-T_2)^2,
					\end{split}
					\\
					\begin{split}
						A\beta_{p_1r}&=-m\gamma\kappa(T_1-T_2)\big[m\kappa T_1+(2\gamma^2+m\kappa)T_2\big],
					\end{split}
					\\
					\begin{split}
						A\beta_{p_2r}&=-m\gamma\kappa(T_1-T_2)\big[(2\gamma^2+m\kappa)T_1+m\kappa T_2\big],
					\end{split}
				\end{align}
			\end{subequations}
			where
			\begin{align}
				\begin{split}
					A&=\frac{1}{2}m(T_1+T_2)\big[m^2\kappa^2(T_1+T_2)^2\\
					&\hspace{1.2in}+4(\gamma^4+2m\gamma^2\kappa)T_1T_2\big].
				\end{split}
			\end{align}
			
		\subsection{\label{subapp:N underdamped particles}Kinetic energy of $N$ identical underdamped particles}
			The Langevin equations are 
			\begin{equation}
				m\ddot{\boldsymbol{r}}_i+\gamma\dot{\boldsymbol{r}}_i+\kappa\sum_{j\ne i}(\boldsymbol{r}_i-\boldsymbol{r}_j)=\sqrt{2T_i\gamma}\boldsymbol{\xi}_i.
			\end{equation}
			Fourier transforming and using $\tilde{\boldsymbol{v}}_i=i\omega\tilde{\boldsymbol{r}}_i$, we have  
			\begin{equation}
				\sum_{j=1}^{N}M_{ij}\tilde{\boldsymbol{v}}_j=i\omega\sqrt{2T_i\gamma}\tilde{\boldsymbol{\xi}}_i,
			\end{equation}
			where
			\begin{equation}
				M_{ij}=(-m\omega^2+i\gamma\omega+N\kappa)\delta_{ij}-\kappa,
			\end{equation}
			the inverse of which is
			\begin{equation}
				(M^{-1})_{ij}=\frac{(-m\omega^2+i\gamma\omega)\delta_{ij}+\kappa}{(-m\omega^2+i\gamma\omega+N\kappa)(-m\omega^2+i\gamma\omega)}.
			\end{equation}
			The power spectrum of the velocities is given by $\langle\tilde{\boldsymbol{v}}_i(\omega)\cdot\tilde{\boldsymbol{v}}_i(\omega')\rangle=2\pi(\boldsymbol{v}_i^2)_{\omega}\delta(\omega+\omega')$ or
			\begin{equation}
				(\boldsymbol{v}_i^2)_{\omega}=\sum_{j=1}^{N}2dT_j\gamma\omega^2(M^{-1})_{ij}(\bar{M}^{-1})_{ij},
			\end{equation}
			where $\bar{M}$ is the complex conjugate of $M$.  By the Weiner-Khinchin theorem, the fluctuations and power spectra are related through
			\begin{equation}
				\langle \boldsymbol{v}_i^2\rangle=\int_{-\infty}^{\infty}\frac{d\omega}{2\pi}\,(\boldsymbol{v}_i^2)_{\omega}.
			\end{equation}
			Performing contour integration over the appropriate poles, we arrive at Eq.\ (\ref{eq:N particle KE}) in the main text.
			
		\subsection{\label{subapp:underdamped one spring}Softening or stiffening one spring, underdamped particles}
			Taking $\kappa_{23}=\kappa_{13}=\kappa\ne\kappa_{12}$, the kinetic energies $\langle K_i\rangle=\left\langle\frac{1}{2}mv_i^2\right\rangle$ in the case of underdamped particles are
			\begin{subequations}
				\begin{align}
					\begin{split}
						\langle K_1\rangle&=\frac{d}{2}T_1-\frac{dm\kappa}{2\gamma^2+3m\kappa}(T_1-T_{\textrm{avg}})\\
						&\hspace{0.8in}+A(\kappa-\kappa_{12})(T_1-T_2),
					\end{split}\\
					\begin{split}
						\langle K_2\rangle&=\frac{d}{2}T_2-\frac{dm\kappa}{2\gamma^2+3m\kappa}(T_2-T_{\textrm{avg}})\\
						&\hspace{0.8in}-A(\kappa-\kappa_{12})(T_1-T_2),
					\end{split}\\
					\langle K_3\rangle&=\frac{d}{2}T_3-\frac{dm\kappa}{2\gamma^2+3m\kappa}(T_3-T_{\textrm{avg}}),
				\end{align}
			\end{subequations}
			where
			\begin{align}
				\begin{split}
					A&=\frac{dm}{12}\left[\frac{8\gamma^2}{(2\gamma^2+3m\kappa)(2\gamma^2+m\kappa+2m\kappa_{12})}\right.\\
					&\hspace{0.5in}\left.-\frac{\kappa-\kappa_{12}}{\kappa(2\gamma^2+m\kappa)+\kappa_{12}(\gamma^2-2m\kappa+m\kappa_{12})}\right].
				\end{split}
			\end{align}
			The average potential energies $\langle U_{ij}\rangle=\left\langle\frac{1}{2}\kappa_{ij}(\boldsymbol{r}_i-\boldsymbol{r}_j)^2\right\rangle$ are
			\begin{subequations}
				\begin{align}
					U_{12}&=\frac{d\kappa_{12}(T_1+T_2)}{2(\kappa+2\kappa_{12})},\\
					\begin{split}
						U_{23}&=\frac{1}{24}\left[T_1+T_2+4T_3+\frac{3\kappa(T_1+T_2)}{\kappa+2\kappa_{12}}\right.\\
						&\hspace{0.4in}\left.-\frac{6\gamma^2\kappa(T_1-T_2)}{\kappa(2\gamma^2+m\kappa)+\kappa_{12}(\gamma^2-2m\kappa+m\kappa_{12})}\right]
					\end{split}\\
					\begin{split}
						U_{13}&=\frac{1}{24}\left[T_1+T_2+4T_3+\frac{3\kappa(T_1+T_2)}{\kappa+2\kappa_{12}}\right.\\
						&\hspace{0.4in}\left.+\frac{6\gamma^2\kappa(T_1-T_2)}{\kappa(2\gamma^2+m\kappa)+\kappa_{12}(\gamma^2-2m\kappa+m\kappa_{12})}\right]
					\end{split}
				\end{align}
			\end{subequations}

\end{document}